\documentclass[12pt]{article}
\usepackage{amsmath,amssymb,amscd,enumerate}
\usepackage{graphics}

\newcommand{\eps}{\varepsilon}
\newcommand{\diff}[2]{\frac{\partial #1}{\partial #2}}
\newcommand{\xxe}{\left(x,\frac{x}{\eps}\right)}
\newcommand{\Ma}{macroscopic }
\newcommand{\mi}{microscopic }

\begin{document}

\title{The Heterogeneous Multi-Scale Method}

\author{Weinan E\footnote{Department of Mathematics and PACM, Princeton
University, and School of Mathematics, Peking University} \and Bjorn
Engquist\footnote{Department of Mathematics and PACM, Princeton
University and Department of Mathematics, University of California, Los
Angeles}}

\maketitle

\begin{abstract}

The heterogeneous multi-scale method (HMM) is a general strategy for
dealing with problems involving multi-scales, with multi-physics, using
multi-grids. It not only unifies several existing multi-scale methods,
but also provide a methodology for designing new algorithms for new
applications. In this paper, we review the history of multi-scale
modeling and simulation that led to the development of HMM, the
methodology itself together with some applications, and the mathematical
theory of stability and accuracy.
\end{abstract}

\section{Introduction}

In the past several years, there has been an explosive growth of
interest on numerical computations for problems involving multi-scales,
often with multiple levels of physical models and use multi-grids. 
Applications of these ideas are found in
many different areas, including coupling quantum mechanics with
molecular
dynamics \cite{CP,Warshel,Yang}, coupling atomistics with continuum
theory \cite{Tadmor1, TimK2, Rudd2, Rob, Cai, EH1, Li, Had}, coupling
kinetic
theory with continuum theory \cite{French, Garcia, Ottinger,
Suen}, coupling kinetic Monte Carlo methods with continuum theory
\cite{Tim}, homogenization theory \cite{Babuska, ER, Hou, Schwab,
Hughes}, and coarse-grained bifurcation analysis \cite{TQK, RTK}.
>From the point of view of numerical analysis, it is natural to ask
whether these seemingly different applications can be put into a common
framework, and whether a general theory of stability and accuracy can be
provided. Such a theory should help us to improve existing methods,
design new ones, and extend their applicability to other problems.

Finite element method provided an example of a success story of this
kind. The practice of finite element methods was started by structural
engineers on very specific applications.  However, the work of
mathematicians at the end of the
60's and early 70's put finite element method on a much more solid and
general framework. This framework provided a thorough understanding of
existing methods, suggested improvements and as well as extension to
other
areas such as fluid mechanics, electromagnetism, etc.

In this article,
we will review the work in \cite{EE} that aimed at providing a general
framework as well as a stability and accuracy theory for
numerical methods involving multiscales, multi-grids and multi-physics.
For convenience and to emphasize the multi-physics nature of these
methods, we will call them Heterogeneous multiscale methods. 
In contrast, typical multi-grid methods
are homogeneous in the sense that they use the same model at different
levels of grids and are aimed at efficiently resolving the details at
the finest grids.

To begin with, let us make some remarks about problems with multiple
scales. Such problems are found everywhere around us. A classical
example that has been extensively studied in the mathematics literature
is the problem of homogenization.
\begin{equation}
\label{eq:1.1}
-\nabla\cdot\left(a\xxe\nabla u^\eps(x)\right)=f(x),\quad x\in\Omega
\end{equation}
with Dirichlet boundary condition $u^\eps|_{\partial\Omega}=0$, for
example \cite{BLP}. Here the multiscale nature is reflected in the
coefficients $a\xxe,\eps\ll1$. In the simplest models
$a(x,y)$ is assumed to be periodic in $y$, say with period
$I=[0,1]^d$, $d$ is the dimension of the physical space. \eqref{eq:1.1}
can be used for modelling transport properties in a medium with
microstructure, and the oscillatory
nature of $a$ is used to model the microstructures in the medium.

Traditionally problems of this type are dealt with either analytically
or empirically by finding effective models that eliminate the small
scales. For the homogenization problem \eqref{eq:1.1}, this means
replacing \eqref{eq:1.1} by a homogenized equation \cite{BLP, Sal} or
effective equation
\begin{equation}
\label{eq:1.2}
-\nabla\cdot(A(x)\nabla U(x))=f(x)\qquad x\in\Omega.
\end{equation}
The solution $U$ of this equation approximates
the behavior of $u^\eps$ averaged over length scales that are much
larger than $\eps$ but smaller than the slow variations of $a$ and $f$.
In the special case when $d=1$, $A(x)$ is given simply by
\begin{equation}
\label{eq:1.3}
A(x)=\left(\int^1_0\frac1{a(x,y)}dy\right)^{-1}
\end{equation}

An althernative to such analytical methods is the empirical modelling.
As an example, let us consider simple incompressible fluids. Let
$u$ be the velocity field. Mass and momentum conservation gives,
\begin{equation}
\label{eq:1.4}
u_t+(u\cdot\nabla)u+\frac1\rho\nabla
p=\nabla\cdot\tau_d,\quad\nabla\cdot u=0
\end{equation}
where $\rho$ is the (constant) density, $\tau_d$ is the viscous stress,
which is a macroscopic idealization of the internal friction forces due
to the short-ranged molecular interactions.
$\tau_d$ must be modeled in order to close the equation \eqref{eq:1.4}.
The simplest empirical model is to assume that $\tau_d$ is linearly
related to $D=\frac{\nabla u+(\nabla u)^T}2$, the rate of strain tensor.
Using homogeneity, isotropy and incompressibility  gives the
constitutive relation
\begin{equation}
\label{eq:1.5}
\tau_d=\nu D
\end{equation}
where $\nu$ is the viscosity of the fluid. In this model, all molecular
details are lumped into a single number $\nu$. It is quite amazing that
such a simple ansatz describes very well the behavior of simple liquids
in almost all regimes.

Despite all these successes, such traditional approaches also have their
limitations. Although a nice set of equations can be derived rigorously
for the effective coefficients $A(x)$ in \eqref{eq:1.2}, they are not
explicit and evaluating them involves a considerable amount of work.
Finding empirical constitutive relations for complex fluids such as
polymeric fluids or liquid crystals has also proven to be a difficult
task. In general the constitutive relations tend to contain many
empirical parameters and their accuracy is also often in serious doubt.

Such problems are not limited to hydrodynamics where constitutive
relations are needed, but is common to most modeling process. In
molecular dynamics, one needs to model, often empirically, the atomic
potentials. In kinetic Monte Carlo methods, one needs to model the
transition rates. In kinetic equations, one needs to model the collision
cross-section. In nonlinear elasticity, one needs to model the
stored-energy functional. In typical mean field theories, one needs to
model the effective local fields and the free energy functional. In
general, the empircal models work well for relatively simple systems,
but lose their accuracy for complex systems.

In the last decade, a new approach, the ``first principle-based''
approach, has been vehemently pursued in various areas of applications.
The basic idea is to replace empirical models by coupling with direct
numerical simulations at a more microscopic level. Some of the best
known examples of this approach include:

1. {\it Ab initio} molecular dynamics \cite{CP}. Here one replaces
the
empirical atomic potential in molecular dynamics by explicit
calculations of the electronic structures. The Car-Parrinello method is
a practical way of implementing this idea \cite{CP}.

2. Quasi-continuum method \cite{Tadmor1,Tadmor2}. This is a method
for
doing nonlinear elasticity calculations without the need of a
stored-energy functional. Instead, the stored energy is computed
directly from atomic potentials using the Cauchy-Born rule. We will
return to this later.

3. The Gas-kinetic scheme \cite{Xu}. This is a method for doing
gas dynamics calculations using directly the kinetic equations instead
of the hydrodynamic equations. Since it played an important role in the
framework developed in \cite{EE}, we briefly review the important steps
here.

Given $\{\rho^n,u^n,T^n\}$, the density, velocity and temperature at
time step $n$ at each cell, the corresponding values at
the next time step, $\{\rho^{n+1},u^{n+1},T^{n+1}\}$ are computed
by:

Step 1. Reconstruction. From $\{\rho^n,u^n,T^n\}$, reconstruct $f^n$,
the one particle phase space distribution function near the cell
boundaries.

Step 2. Solve the kinetic equation with initial data $f^n$ near the
cell boundaries. In \cite{Xu}, the kinetic equaiton is chosen as
the BGK equation
$$f_t+(u\cdot\nabla)f=\frac1\eps(\mathcal{X}_{(\rho,u,T)}-f)$$
where $\mathcal{X}_{(\rho,u,T)}$ is the local Maxwellian associated with
$(\rho,u,T)$.

Step 3. Compute the average density, momentum and energy fluxes at the
cell boundaries, from which one computes
$\{\rho^{n+1},u^{n+1},T^{n+1}\}$.

This procedure is an illustration of several of the key ingredients that
we use in the general framework that we will discuss below: 
the selection of an overall macroscale scheme which in the present
example is the finite volume method; the estimation of the
macroscale data, here the flux, using the Godunov
procedure which consists of the steps of reconstruction, microscale
evolution and
averaging; the cost reduction  at the microscale 
evolution step by restricting to a small subset of the
computational domain.


The examples we discussed so far belong to the class of problems
referred to as type B problems in \cite{EE} where microscopic models are
used to supply a closure to the macroscopic models. Another wide class
of problems, called type A problems in \cite{EE} consist of problems
with defects, interfaces or singularities, for which conventional
macroscopic models are accurate enough away from the defects, and more
detailed microscopic models are necessary near the defects. Type A
problems are found in crack propagation, contact line dynamics, triple
junctions, grain boundary motion, dislocation dynamics, etc.

This short review of existing work is by no means comprehensive.
For the convenience of the reader, 
we include at the end an extensive list of references.

\section{Relations between Macroscopic and Microscopic Models}

Let us first fix the notations. We will denote the \Ma and \mi state
variables as $U$ and $u$, defined on $D$ and $\mathcal{D}$,
respectively. Typically $D$ is the physical space. As we will explain
below using examples, it is convenient to view $\mathcal{D}$ loosely as
a fiber bundle over $D$, where the fiber $\mathcal{D}_x$ over $x\in D$,
is roughly the space of microstructures at $x$. The \Ma and \mi state
variables are connected by a compression operator $Q$, and a
reconstruction operator $R$
$$Qu=U,\quad RU=u,\quad QR=I$$
where $I$ is the identity operator. Typically there is a natural way to
define $Q$. But $R$ is certainly not unique.

To illustrate these notions and notations, let us consider a few
examples.

\begin{enumerate}
\item For the first example we will consider the case when the \Ma
variables are the hydrodynamic variables of density, velocity and
temperature $(\rho,u,T)$, and the \mi variable is the one particle phase
space distribution function $f$. In this case $D$ is the physical
space-time domain of interest, $\mathcal{D}=D\times R^3$, $\mathcal{D}_x
=R^3$, the momentum space which represents the microstructures in this
problem. $Q$ is defined by
$$\rho(x,t)=\int_{R^3}f(v,x,t)dv,\quad u(x,t)=\int_{R^3}f(v,x,t)
vdv,\quad T(x,t)=\frac1{3\rho}\int_{R^3}f(v,x,t)|v-u|^2dv$$
\item For our second example, we will take the \mi model to be the
kinetic models of rod-like molecules in liquid crystals, \cite{Doi}
where the \mi variable is the orientation-position distribution function
$f(x,m,t)$, $(x,t)\in D$, $m\in S^2$, the unit sphere. Here
$\mathcal{D}=D\times S^2$, $\mathcal{D}_x=S^2$. The \Ma model is the
Landau-de Gennes type models with tensorial order parameter $S$ which is
our \Ma variable. $Q$ is defined as
$$S(x,t)=\int_{S^2}\left(m\otimes m-\frac13I\right)f(x,m,t)dm$$
\item For the third example, we take the standard homogenization problem
$$u^\eps_t+\left(a\xxe u^\eps\right)_x=0$$
where $a(x,y)$ is smooth and periodic in $y$ with period 1. In this
case, the \Ma variable will be the local space-time averages of
$u^\eps$:
$$U(x,t)=\frac1{|C|}\int_{{C}}u^\eps(x+y,t+s)dyds$$
where $|C|$ denotes the area of $c$ on which the averaging is taken.
$D_x=(x,t)+C$. The size of $C$ should be larger than $\eps$.
\item Our last example is front propagation described by Ginzburg-Landau
equations
$$u^\eps_t=\Delta u^\eps+\frac1{\eps^2}u^\eps(1-(u^\eps)^2)$$
To define the \Ma variables, observe that $u^\eps$ is close to $\pm1$ in
most of the physical domain $D$, except in a thin region of thickness
$O(\eps)$ where sharp transition between $\pm1$ takes place. Since the
fast reaction term vanishes at three values $-1,0,+1$, it is natural to
define $U$ by:
$$U(x,t)=Qu^\eps(x,t)=\left\{\begin{array}{rl}
1 & \mbox{if }u^\eps(x,t)>0 \\
0 & \mbox{if }u^\eps(x,t)=0 \\
-1 & \mbox{if }u^\eps(x,t)<0
\end{array}\right.$$
An equivalent definition of $U$ is via the {\it 0} level set of
$u^\eps$.

More examples are disscussed in \cite{EE}. 
\end{enumerate}

In the following we will concentrate on problems of type B, namely
problems for which there exist a set of macroscopic variables that obey
closed \Ma models, but the \Ma models are not explicitly available. We
will describe general computational methodologies that enable us to do
numerical computations efficiently based on the underlying \mi models.
We will comment on problems of type A at the end.

\section{Abstract Formulations}

A key component of HMM is the estimation of macroscale data using the
microscale model. 
To see how this can be done, it is helpful to first 
 give an abstract formulation of the macroscopic
model in terms of the microscopic model. We start with variational
problems.

Consider a microscopic variational problem
\begin{equation}
\label{eq:3.1}
\min_{u \in \omega} e(u)
\end{equation}
where $\omega$ is some function space over $\mathcal{D}$, the physical
space.
Let $Q$ be the compression operator. $Q$ maps $\omega$ to $\Omega$, a
function space over $D$.  Since
\begin{equation}
\label{eq:3.2}
\min_{u\in\omega}e(u)=\min_{U\in\Omega}\min_{Qu=U}e(u),
\end{equation}
our \Ma variational problem is given by
\begin{equation}
\label{eq:3.3}
\min_{U\in\Omega}E(U)
\end{equation}
where
\begin{equation}
\label{eq:3.4}
E(U)=\min_{Qu=U}e(u)
\end{equation}

For dynamic problems, let us denote by $s(t)$, the evolution operator
for the \mi process. In general $\{s(t),t>0\}$ forms a semi-group of
operators. This semi-group may be generated by a set of differential
equations, a Markov process, or a discrete dynamical system. There are
at two important time scales in our problem: $t_R$, the relaxation time
scale of the \mi process, and $t_M$, the \Ma time scale of interest. Our
basic assumption that there exists a well-defined \Ma model over the \Ma
time scale of interest implies that $t_R\ll t_M$.

Denote by $R$ an appropriately chosen reconstruction operator. Given
$U\in\Omega$, let 
\begin{equation}
\label{eq:3.5}
S(t)U=Qs(t)RU
\end{equation}
Obviously $S(t)U$ depends on $R$ as it is defined. However, since
$t_R\ll t_M$, we expect that the dependence on $R$ diminishes for $t\gg
t_R$. Therefore we can define the \Ma model approximately as
\begin{equation}
\label{eq:3.6}
U_t=F(U)
\end{equation}
where
$$F(U)=\frac{S(\triangle t)U-U}{\triangle t}$$
with appropriately chosen $\triangle t$, $t_R\ll\triangle t\ll t_M$.

\section{The Structure of HMM}

Our basic numerical strategy is now as follows. We will work with a \Ma
grid that adequately resolve the \Ma problem, but do not necessarily
resolve the \mi problem, and we will attempt to solve directly the \Ma
model \eqref{eq:3.6} and \eqref{eq:3.3}. 

There are two  main components
in the heterogeneous multiscale method: {\it An overall macroscopic
scheme
for $U$ and
estimating the missing macroscopic data from the microscopic model}.

\subsection{The Overall Macroscopic Scheme}

The right overall macroscopic scheme depends on the nature of the problem
and typically there are more than one choice. For variational problems,
we can use the standard finite element method. In fact our examples
in the next section use the standard piecewise linear finite element
method. For dynamic problems that are conservative, we may use the
methods developed for nonlinear conservation laws (see, e.g. \cite{LeVeque}).
Examples include the Godunov scheme, Lax-Friedrichs scheme, and the
discontinuous Galerkin method.
For dynamic problems that are non-conservative, one could simply
use a standard ODE solver, such as the forward Euler or the Runge-Kutta
method, coupled with the force estimator that we discuss below.


\subsection{Estimation of the Macroscopic Data}
After selecting the overall macroscopic scheme, we face the
difficulty that not all data for the macro scheme are available since
the underlying macro model is not explicitly known. The next component
of HMM is to estimate such missing data from the microscopic model. This
is done by solving the micro model locally subject to the constraint
that
$\tilde{Q} u = U$ where $\tilde{Q} $ is the approximation to $Q$ and
$U$ is the current macro state. For example, for the variational
homogenization problem, the missing data is the
stiffness matrix for the macro model. As we explain in the next section,
this data can be estimated by solving the original microscopic
variational homogenization problem on a unit cell in each element of the
triangulation, subject to the constraint that $\tilde{Q} u = U$.
For dynamic problems, such data can be estimated from a Godunov
procedure, namely, that we first reconstruct the micro state from $U$,
and evolve the micro state subject to the constraint that $\tilde{Q} u =
U$, and then estimate the missing data from $u$. The missing data can
be either the forces or fluxes, or a part of the forces or fluxes
such as the eddy viscosity term in a turbulence model. We also have the
option of carrying out a number of such microscopic calculations
(e.g. with different
reconstruction or different realization of the randomness) and extract
a more accurate estimate from the collection of microscopic
calculations.

\subsection{Examples}
To illustrate the selection of the macroscale scheme and the estimation
of missing macroscale data from microscale models, we will discuss some
examples in more detail.

{\bf 1.  Variational Problems}

Examples include
\begin{enumerate}
\item $$\min_{u\in
H^1_0(D)}\int_D\left\{\frac12\sum_{i,j}a^\eps_{i,j}(x,u)\diff{u}{x_i}\diff{u}{x_j}
-f(x)u(x)\right\}dx$$
where the multiscale nature of the problem is contained in the tensor
$a^\eps(x,u)=(a^\eps_{i,j}(x,u))$ which can be of the form
\begin{enumerate}
\item $a^\eps(x,u)=a\left(x,\frac{x}\eps\right)$, where $a(x,y)$ is
smooth and periodic in $y$ with period $[0,1]^d$. This is the classical
homogenization problem we discussed earlier.
\item $a^\eps(x,u)=a\left(x,\frac{x}\eps\right)$, where $a(x,y)$ is
random and stationary in $y$. This can be used to model random medium.
\item $a^\eps(x,u)=a\left(x,u,\frac{x}\eps\right)$, where $a(x,u,y)$ is
smooth. The dependence on $u$ makes this problem nonlinear. The
dependence on $y$ can be either periodic or random stationary.
\end{enumerate}
The macroscale problem is of the type
$$\min_{U\in H^1_0(D)}\int_D\left\{
\frac12A(x,U,\nabla U)-f(x)U(x)\right\}dx$$
\item Atomistic models of crystalline solids:
$$\min_{\{x_j\}} \sum_{y_i,y_j\in
D}V(x_i-x_j)$$
subject to loading conditions, where $V$ is a pairwise
atomistic potential, $x_i=y_i+u_i$,$y_i$ is the position of the $i$-th
atom
before deformation, $u_i$ is the displacement of the
$i$-th atom. The macroscale problem is of the type considered in
nonlinear elasticity
$$\min_U \int_Df(\nabla U)$$
where $U$ is the macroscale displacement field.
\end{enumerate}
For these problems, we can choose the macroscale scheme to be the
standard finite element method over a macroscale triangulation. The
macroscale data that we need to estimate is either $\int_DA(x,U,\nabla
U)dx$ or $\int_Df(\nabla U)dx$ for $U\in V_H$, the finite element
space. These can be approximated via the following steps.
\begin{enumerate}
\item For each element $\Delta$, approximate $\int_\Delta A(x,U,\nabla
U)dx$ or $\int_\Delta (\nabla U)dx$ by a quadrature formula.
\item For each quadrature nodes
$x_i\in\Delta$,
approximate $A(x,U,\nabla U)(x_i)$ or
$f(\nabla U)(x_i)$ by minimizing the original microscale
problem over a micro-cell $\Delta_{x_i}$, subject to the constraint that
$\int_{\Delta_{x_i}}u(x)dx=\int_{\Delta_{x_i}}U(x)dx,\int_{\Delta_{x_i}}\nabla
u(x)dx=\int_{\Delta_{x_i}}\nabla U(x)dx$, with appropriate changes for
the atomistic problem. For the periodic homogenization and crystalline
solids problems, $\Delta_{x_i}$ can be chosen to be a unit cell around
$x_i$, if we replace the constraint by a periodic boundary condition or
the Cauchy-Born rule, as we explain in the next section. For the
stochastic homogenization problem, $\Delta_{x_i}$ should be larger than
the correlation length. In this case, it may also be of advantageous to
perform ensemble averages over several realizations of
$a\left(x,\frac{x}\eps\right)$.
\end{enumerate}

{\bf 2. Dynamic Problems of Conservative Type}

Examples include
\begin{enumerate}
\item $$\partial_tu^\eps=\nabla\cdot(a^\eps(x,u)\nabla u^\eps)$$
 where
$\{a^\eps(x,u)\}$ is as discussed above.
\item $$\partial_tu^\eps+\nabla\cdot(a^\eps(x)u)=0$$
 where
$a^\eps(x)=a\left(x,\frac{x}\eps\right), a(x,y)$ can either be periodic
 or stochastic stationary in $y$.
\item Kinetic models such as the Boltzmann or BGK equations.
\item Molecular dynamics of the type discussed in Section 2.
\item Spin-exchange models via Kawasaki dynamics \cite{Spohn}.
\end{enumerate}
Other examples may include models of phase segregation, mixtures of
binary fluids, elastic effects, etc. The macroscale models are of the
type
\begin{equation}
\label{eq:macro}
U_t+\nabla\cdot J=0
\end{equation}
where $U$ is in general a vectorial macroscale variable, $J$ may depend
on $x,U,\nabla U$, etc.

The macroscale scheme can be either a finite volume method, such as the
Godunov scheme, or a finite element method, such as the discontinuous
Galerkin method. We will discuss here the finite volume method. HMM
based on the discontinuous Galerkin method is considered in \cite{EES}.

The missing macroscale data for a finite volume method for
\eqref{eq:macro} is the
macroscale flux $J$ at the cell boundaries denoted by
$\{J_{j+\frac12}\}$. They can be estimated by the following
``Godunov-like'' procedure
\begin{enumerate}
\item Select a microcell $\Delta_{j+\frac12}$ around the cell boundary
at $x_{j+\frac12}$.
\item From $\{U^n_j\}$, reconstruct the microstates $\{\tilde{u}\}$ on
$\{\Delta_{j+\frac12}\}$. $\tilde{u}$ should be consistent with
$\{U^n_j\}$ in the sense that $\tilde{Q}u=U^n$, where $\tilde{Q}$ is the
approximation of $Q$ restricted to $\{\Delta_{j+\frac12}\}$.
\item Evolve the microstate $u(t)$ using the microscale model inside
$\{\Delta_{j+\frac12}\}$, with initial state $\{\tilde{u}\}$, and
subject to the constraint that
$$\tilde{Q}u(t)=U$$
\item Evaluate the macroscale flux $\{J_{j+\frac12}\}$ using $\{u(t)\}$.
\end{enumerate}
The constraint $\tilde{Q}u=U$ requires some additional comment. Take the
example of molecular dynamics. If we would like to capture the
macroscale behavior at the level of Euler's equations, the constraint is
simply that the average  mass, momentum and energy should be given by
the prescribed macroscale values given by $\{U^n\}$. If we would like to
capture the viscous or higher order effects, we also need to constrain
the system such that the average density, momentum and energy gradients
be given by the macroscale values. This is less convenient to implement,
particularly if higher order gradients are required. The discontinuous
Galerkin formulation proposed in \cite{EES} avoids this difficulty.

The rules for selecting $\{\Delta_{j+\frac12}\}$ is the same as for the
variational problems. As usual for periodic homogenization and
crystalline solids problems, $\Delta_{j+\frac12}$ can be chosen to be
the unit cell.

{\bf 3.  Dynamic Problems of Nonconservative Type}

Examples include
\begin{enumerate}
\item
$$\partial_tu^\eps=\sum_{i,j}a^\eps_{i,j}(x,u)
\frac{\partial^2u}{\partial x_i\partial x_j}$$
 where $\{a^\eps(x,u)\}$ is as discussed before.
\item Spin flip models \cite{Joel} that leads to Ginzburg-Landau type
of equations.
\end{enumerate}

In this case, we write the macroscale model as
$$U_t=F(U)$$
where $F(U)$ can be a nonlinear operator acting on $U$. For the
macroscale scheme, we choose an ODE solver on a grid, such as forward
Euler or Runge-Kutta, and we need to estimate $F(U)$ on the macro grid.

For each macro grid point $x_j$, we again select a microcell $\Delta_j$
around $x_j$. The principle for selecting $\Delta_j$ is the same as
before. From $\{U^n_j\}$, we construct a piecewise polynomial of $k$-th
order in $\Delta_j$ denoted by $U^n_j(x)$. The rest of the steps are the
same as that for the conservative systems. We note that the constraint
$\tilde{Q}u=U$ can be interpreted as
$$\int_{\Delta_j}(u(x)-U^n_j(x))x^mdx=0$$
for $0\le m\le k$.

{\bf 4. Macroscale Markov Chains}

When the macroscale process is a Markov chain, it is natural to
use a kinetic Monte Carlo method as the macroscale scheme. The missing
data
might be the transition rates between macro states. Estimating such data
is a rather non-trivial task. It is discussed in \cite{ERV}.

In the following, for dynamic problems we will concentrate on the
simplest
case when the macroscopic scheme itself is a Godunov scheme and the
missing data
is the macroscopic forces. Extension to more general situations will be
studied in \cite{EE2}.

\section{Compression Techniques}

The key numerical problem now is how to construct approximations to
$E(U)$ and $F(U)$. In this section, we review numerical techniques for
efficiently approximating $E(U)$ and $F(U)$ by exploiting the separation
of spatial/temporal scales.

\subsection{Compression in the Spatial Domain}
 If the \Ma and \mi
spatial scales are separated, we can effectively reduce the
approximation of $E(U)$ and $F(U)$ to a unit of \mi size on each \Ma
cell. Such an idea is embodied in e.g. the quasi-continuum method. We
will illustrate this with some examples.

Example 1. The Variational Homogenization Problem

Consider the variational problem
\begin{equation}
\label{eq:4.1}
\min_{u\in
H^1_0(D)}\frac12\sum_{i,j}\int_Da_{ij}\left(x,\frac{x}\eps\right)
\diff{u}{x_i}\diff{u}{x_j} dx-\int_Df(x)u(x)dx
\end{equation}
where as usual we assume that $a(x,y)$ is smooth and periodic in $y$
with period $I=[0,1]^d$. Let $T_H$ be a \Ma triangulation of $D$
and $V_H\in H^1_0(D)$ be the standard piecewise linear finite
element space over $T_H$. For $u\in H^1_0(D)=\omega$, define $Qu=U\in
V_H=\Omega$, if
\begin{equation}
\label{eq:4.2}
\int_K\nabla udx=\int_K\nabla Udx
\end{equation}
for all triangles $K\in T_H$. $E(U)$ as defined in
\eqref{eq:3.3} involves nonlocal coupling of all the triangles. However,
we can approximate $E(U)$ efficiently if $\eps\ll1$. This is done as
follows. 
Given $U\in V_H$ and $K\in T_H$, denote
by $x_K$ the center of mass of $K$, and $u_K$
the
solution of the problem
\begin{equation}
\label{eq:4.3}
\min\frac12\int_{x_K+\eps
I}\sum_{i,j}a_{ij}\xxe\diff{u}{x_i}\diff{u}{x_j}dx
\end{equation}
subject to the condition
\begin{equation}
\label{eq:4.4}
u(x)-U(x)\mbox{ is periodic with period }\eps I.
\end{equation} 
Let
\begin{equation}
\label{eq:4.5}
A_K(U,U)=\frac{|K|}{\eps^d}\int_{x_K+\eps
I}\sum_{i,j}a_{ij}\xxe\diff{u_K}{x_i}\diff{u_K}{x_j}dx
\end{equation}
where $|K|$ is the volume of $K$, 
we then approximate $E(U)$ by
\begin{equation}
\label{eq:4.6}
\tilde{E}(U)=\frac12\sum_K A_K(U,U)-\int_DU(x)f(x)dx
\end{equation}
In this example, the computation on the microscale model is reduced to a
microscopic unit-cell problem on each \Ma element.

The complexity of this method is comparable to solving directly the
homogenized equation by evaluating the coefficients of the  homogenized
equations on each element. This is the minimal one can hope for.
However, our method differs from solving the homogenized equation in one
essential aspect: Our method is based on the finite $\eps$-microscale
model, not the homogenized equation which represents the $\eps\to0$
limit. Consequently our method can be readily extended to more complex
problem such as the nonlinear homogenization problem at essentially the
same cost.

Example 2. Quasi-Continuum Method \cite{Tadmor1, Tadmor2}.

This is a way of doing nonlinear elasticity calculations using only
atomic potentials. Denote by $x_1,x_2,\ldots x_N$ the positions of all
the individual atoms in a crystal. At zero tempuerature, the position of
the atoms are determined by
\begin{equation}
\label{eq:4.7}
\min\left\{V(x_1,x_2,\ldots x_N)-\sum^N_{i=1}f(x_i)u_i\right\}
\end{equation}
subject to appropriate boundary conditions. Here $f$ is the external
force, $u_i$ is the displacement of the $i$-th atom, $V$ is the
interaction potential between the atoms.

Quasi-continuum method works with a \Ma triangulation of the crystal and
a standard piecewise linear finite element space for the displacement
field. The compression opeartor is defined in a similar way as in
\eqref{eq:4.2}: $Q_u=U\in V_H$ if
\begin{equation}
\label{eq:4.8}
<\nabla u>_K=\frac1{|K|}\int_K\nabla Udx
\end{equation}
for all $K\in T_H$, where $<\nabla u>_K$ denotes
the average strain of the atoms on the element $K$. Having defined
$Q, E(U)$ is defined as in Section 3.

To approximate $E(U)$, Tadmor et.al. uses the Cauchy-Born rule. Given
$U\in V_H$, let $e_K(U)$ be the potential energy of a unit cell of the
crystal subject to the uniform strain $\nabla U$ on $K$. Let
\begin{equation}
\label{eq:4.9}
\tilde{E}(U)=\sum_K n_K e_K(U)-\int_Df(x)U(x)dx
\end{equation}
where $n_K$ is the number of unit cells on $K$.
$\tilde{E}(U)$
is the approximation of $E(U)$.

Quasi-continuum method contains an additional element for dealing with
defects and interfaces in crystals, i.e. type A problems, by replacing
the Cauchy-Born rule with a full-atom calculation on elements near the
defects and interfaces. We will return to this later.

\subsection{Compression in the temporal domain}

By resorting to the microscopic model in order to simulate
the macroscopic dynamics, we are forced to resolve the microscopic
times scales which are not of interest.  This is particularly
expensive if $t_R\ll t_M$.
However in this case we can
explore this time scale separation  to reduce the computational
cost in the temporal domain.

It is helpful to distinguish two different scenarios by which
relaxation to local equilibrium takes place. For some problems, such as
the
parabolic homogenization problem \eqref{eq:57a}
and  the Boltzmann equation, we have strong convergence to equilibrium.
No temporal or ensemble averaging is necessary for the convergence
of the physical observables.
 For other problems, such as the
advection homogenization problem and molecular dynamics,
convergence to equilibrium is in the sense of distributions, i.e.
physical observables converge to their local equilibrium values
after time or ensemble averaging.
Let us  express the approximate $F(U)$, called a
$F$-estimator,  in  the form
$$\tilde{F}(U)=Q\sum^k_{j=1}\psi_jf(u_j)$$
where the weights $\{\psi_j\}$ should satisfy
$$\sum^k_{j=1}\psi_j=1$$
$u_j$ is the computed microscopic state at microscopic time step $j$,
and $u_0 = RU$ where $R$ is some reconstruction operator.
The selection of  the weights in the $F$-estimator crucially
depends on the nature of this convergence.
In particular we note two special choices.
The first is: $\psi_k=1$ and $\psi_j=0$
for $j<k$. This is suitable when we have strong convergence to
equilibrium. The second choice is: $\psi_j=\frac1k$, for $1\le j\le k$.
This is more suited for the case when we have weak convergence to local
equilibrium.
More accurate choices of the weights are discussed in \cite{EE2}.
The time interval on which the microscopic model has to be solved
depends on how fast the transient introduced by the reconstruction step
dies out. 

Consider the parabolic
homogenization problem
\begin{equation}
\label{eq:57a}
u^\eps_t=\nabla\cdot(a\left(x,\frac{x}\eps\right)\nabla u^\eps)
\end{equation}
on $D$, with Dirichlet boundary condition $u^\eps|_{\partial D}=0$.
To approximate the macroscopic behavior of $u^\eps$, we will work with
a macroscopic grid of size $(\Delta x, \Delta t)$.
Let
$U=Qu^\eps$ be the moving cell averages of $u^\eps$ over a cell of size
$\Delta x$.
Let $R$ be the piecewise linear
reconstruction. In one-dimension, this is
$RU(x)=U_j+\frac{U_{j+1}-U_j}{\Delta x}(x-x_j)$, for $x\in[j\Delta
x,(j+1)\Delta x]$. With this reconstruction, we proceed with the
microscopic solver. Asymptotic analysis suggests that the relaxation
time for this problem is $O(\eps^2)$ \cite{BLP}. We plot in Figure 1 a
typical
behavior of the microscopic flux
$j^\eps(x,t)=a\left(x,\frac{x}\eps\right)\nabla u^\eps(x,t)$ at a cell
boundary over the time interval $[t^n,t^n+\Delta t]$ as a function of
the
micro time steps. It is quite clear that
$j^\eps(x,t)$ quickly settles down (after about 35 micro time steps)
to a quasi-stationary value after a rapid transient.
We obtain an efficient numerical scheme if we select this value as
the macroscopic flux and use that to evolve $U$ over a much larger
time step $\Delta t$.
\begin{center}
\resizebox{3in}{!}{\includegraphics{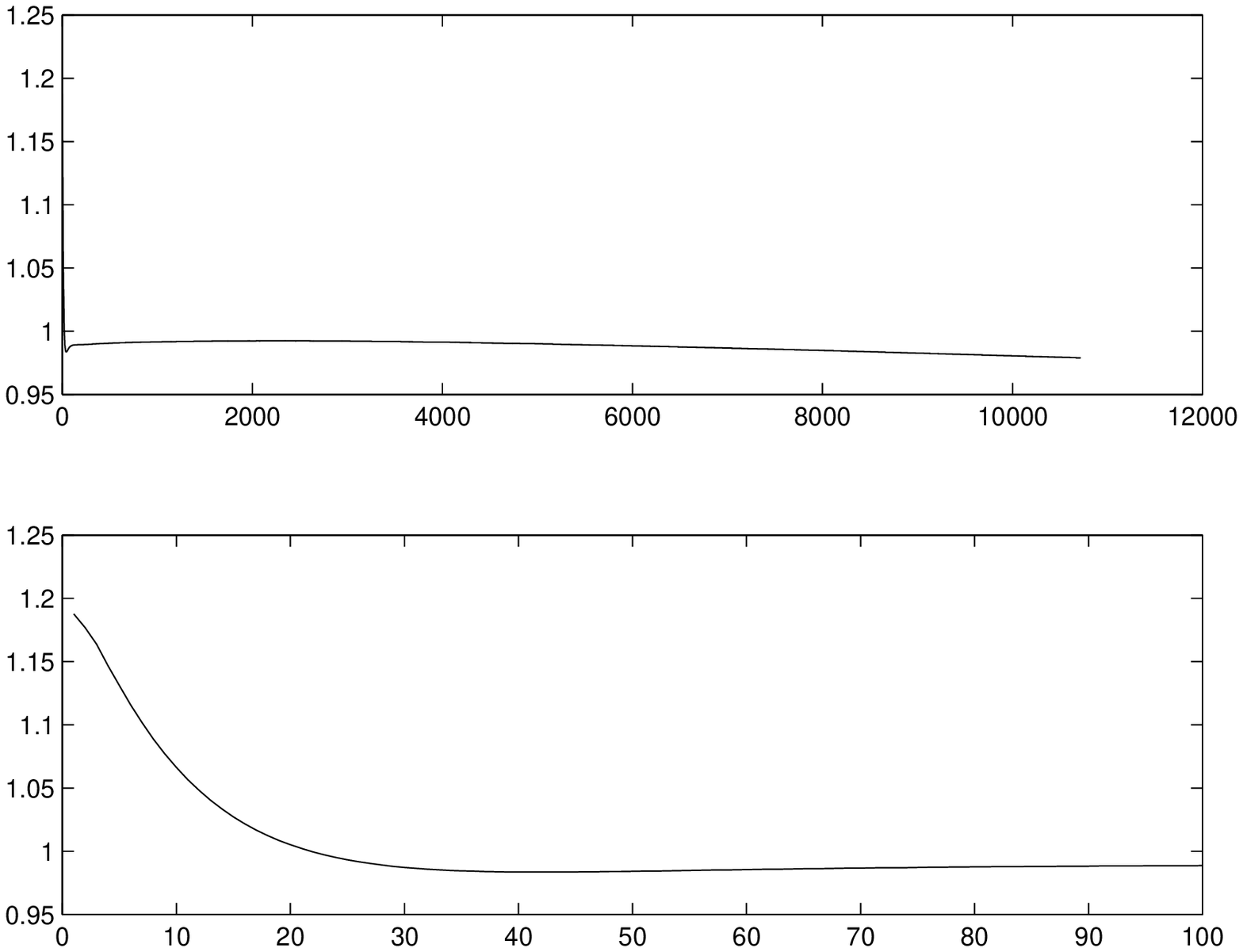}} \\
\end{center}
{\small Figure 1. Computed flux
$\tau^\eps(x,t)=a\left(x,\frac{x}\eps\right)
\nabla u^\eps(x,t)$ as a function of the micro time step over one
typical macro time step, for the parabolic homogenization
$a\left(x,\frac{x}\eps\right)=2+\sin 2\pi\frac{x}\eps$. The bottom
figure is a detailed view of the top figure for small time steps.
Notice that $j^\eps(x,t)$ quickly settles down (after about 35 micro
time steps)
to a quasi-stationary value after a rapid transient.  }

Our next example is the advection homogenization problem
\begin{equation}
\label{eq:58a}
u^\eps_t+\nabla\cdot(a\left(x,\frac{x}\eps\right)u^\eps)=0
\end{equation}
in one-dimension. We assume $a(x,y)>a_0>0$. We proceed as before,
except that we take a piecewise constant reconstruction. In contrast to
the
previous example, the temporal oscillations in the solutions of
\eqref{eq:58a}
do not die out. This is reflected in Figure 2 where we plot the
microscopic flux $j^\eps(x,t)=a\left(x,\frac{x}\eps\right)u^\eps(x,t)$
over the time interval $[t^n,t^n+\Delta t]$ as a function of the
microscale
time steps. $j^\eps$ remains
oscillatory throughout the time interval. Nevertheless, if we plot the
 time average
\begin{equation}
\bar{j}(x,t)=\frac1t\int^{t^n+t}_{t^n}
 K\left(
1-\frac\tau{t}\right)j^\eps(x,\tau)d\tau,
\quad \quad
K(\tau)=1-\cos 2\pi\tau
\end{equation}
as shown in Figure 3, we see that it settles down to a quasi-stationary
value on a time scale of $O(\eps)$.

\begin{center}
\resizebox{3in}{!}{\includegraphics{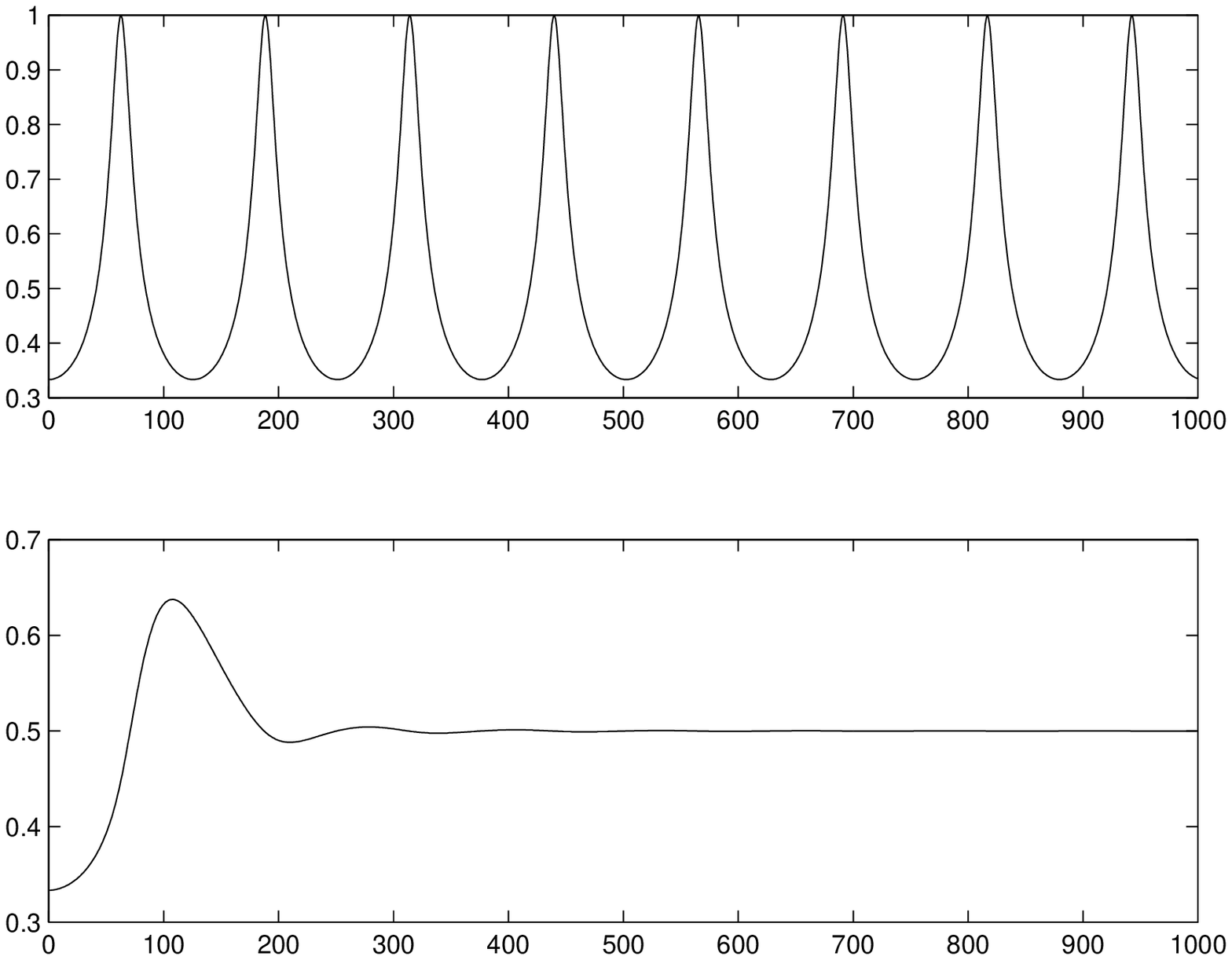}} \\
\end{center}
{\small Figure 2. Top figure: Computed flux
$j^\eps(x,t)=a\left(x,\frac{x}\eps\right)u^\eps$ as a fraction of the
micro time step over one macro time step for the convection
homogenization problem \eqref{eq:58a}. Bottom figure: Time averaged flux
$\bar{j}(x,t)$ as a function of the micro time step.}

The fact that the microscopic process only has to be evolved on
time scales comparable to $t_R$ leads to other possibilities of
state space compression by neglecting the part of the state
space which does not contribute significantly to the
$F$-estimator.

In summary, we can express the $F$-estimator at time $t$ as
\begin{equation}
F^\eps(U,t)=\tilde{Q}_{\Delta t}\{f(\tilde{u}(\tau)),t\le\tau\le t
+\Delta t,\tilde{u}(t)=RU\}
\end{equation}
where $\tilde{u}(t)$ is the solution of the compressed microscopic model
(possibly over a truncated computational domain) with initial data
$\tilde{u}(t)=RU,\tilde{Q}_{\Delta t}$ is the numerical approximation of
the compression operator. Typically $\tilde{Q}_{\Delta t}$ has the form
\begin{equation}
\tilde{Q}_{\Delta t}=Q_eQ_xQ_t
\end{equation}
where $Q_e,Q_x,Q_t$ denote the compression operators over the
probability, spatial and temporal spaces respectively. Having
$F^\eps(U,t)$, the macroscopic state variables can be updated using
standard ODE solvers. The simplest example of forward Euler scheme gives
\begin{equation}
U^{n+1}=U^n+\Delta tF^\eps(U^n,t^n).
\end{equation}

\section{Stability and Accuracy of HMM}

\subsection{Variational Problems}

The analysis of HMM proceeds in the same way as the analysis of
traditional numerical methods, except we have to deal with in addition
the effect of compression. For variational problems, compression gives
rise to additional error in the evaluation of the macroscale energy
functional, or for linear problems, the stiffness matrix.

Take the example of the variational homogenization problem. The main
error in the evaluation of the stiffness matrix comes from the
inconsistency at the surface of the element where it meets another
element. This error is of the order $\frac\eps{\triangle x}\|\nabla
U \|^2_{L^2}$. Consequently, one has

{\bf Theorem 1.} Assume that the finite element triangulation is
quasi-regular, then
$$\|U-Qu^\eps\|_{H^1(D)}\le C\left( H +\frac\eps{H}\right)$$
where $U$ is the numerical solution of HMM, $H$ is the size of the
macroscale
element.

\subsection{Dynamical Problems}

For dynamic problems, it is helpful to define an auxiliary macroscale
scheme, called the Generalized Godunov Scheme (GGS) in \cite{EE}.
Roughly speaking, GGS is obtained if in HMM we replace the microscale
solver in the data estimation step by the macroscale solver. Since the
macroscale model is not explicitly known, the GGS is not a practical
tool but only an analytical tool that is helpful for analyzing HMM.

For example, for the parabolic and advection homogenization problems
discussed earlier, GGS is simply the Godunov scheme on the macroscale
problem with appropriate reconstruction and approximate Riemann solvers.

Assuming that the macroscale model is in the form of a differential
equation $U_t=F(U)$, we can write a one-step HMM in the form
$$U^{n+1}_j=U^n_j+\triangle tF_j(U^n)$$
and GGS in the form
$$\bar{U}^{n+1}_j=\bar{U}^n_j+\triangle t\bar{F}_j(\bar{U}^n)$$
The basic stability result proved in \cite{EE} is that if GGS is stable,
then the HMM is stable and 
$$\|U^n-\bar{U}^n\|\le C(\|U^0-\bar{U}^{-0}\|+\max_{0\le
k\le\frac{T}{\triangle t}}\|\bar{F}(U^k)-F(U^k)\|)$$
for $n\triangle t\le T$.

The notion of stability for the GGS has to be quantified appropriately
for nonlinear problems. See \cite{EE} for details.

Noting that
$$\|U^n-Qu^\eps\|\le\|U^n-\bar{U}^n\|+\|\bar{U}^n-Q u^\eps\|$$
we now conclude that the stability and accuracy of HMM depends on

\begin{enumerate}
\item Consistency of GGS with the macroscale model.
\item Stability of GGS.
\item The compression error $\|\bar{F}(U)-F(U)\|$
\end{enumerate}

We discuss each of these in some more detail.

\begin{enumerate}
\item Consistency of GGS and the macroscale model might be lost if the
overall
macroscale scheme does not probe the macroscale properties to the right
level of accuracy. For example, if the macroscale model is hydrodynamics
including viscous effect, and in HMM we have only  probed the flux in
the convective terms by using a piecewise constant reconstruction near
the cell boundaries,
neglecting the dissipative terms. This results in inconsistency with the
macroscale model. Other such examples are discussed in \cite{EE}.

\item Stability of GGS usually results in the standard constraint on
macro time step size. It may also impose constraints on the
reconstruction operator.

\item The compression error has also several sources, e.g. compressions
in the temporal or spatial domains. The nature of the temporal
compression error depends on the nature of the relaxation to local
equilibrium of the microscale process. In the case of strong relaxation,
no temporal averaging is necessary for macroscale data estimation.  In
the
case of weak relaxation, the temporal compression error depends strongly
on the temporal and/or ensemble averaging operator used.
\end{enumerate}

\cite{EE} also pointed out the importance of averaging out spatial small
scales for HMM based on the flux-formulation.

\section{Conclusion}

There are two important questions that have to addressed in order
to design efficient numerical methods that couple the macro and
microscale models:

\begin{enumerate}
\item What is the best way to set up the individual microscale problems?

\item How do we couple the microscale problems together in order to
simulate the macroscale behavior?
\end{enumerate}

The second question is now fairly adequately addressed by HMM. The first
question is tied more with specific applications. We have discussed a
few examples. But much more work needs to be done in order to understand
the issue in the general case.


{\bf Acknowledgement.}
We are grateful for many inspiring discussions with Yannis Kevrekidis in
which he has outlined his program of macroscale analysis based on
microscale solvers. 
We are also grateful to Eric Vanden-Eijnden and Olof Runborg for
stimulating discussions, and to  Assyr Abdulle and Chris Schwab for
suggestions
that improved the first draft of the paper.
The work of E is supported in part by an ONR grant N00014-01-1-0674.
The work of Engquist is supported in part by NSF grant DMS-9973341.

\end{document}